\documentclass[12pt,preprint]{aastex}
\usepackage{apjfonts}
\usepackage{natbib}
\usepackage{amssymb, amsmath, epsfig, epsf, lscape}
\newcommand{\flux}{erg~s$^{-1}$~cm$^{-2}$}
\newcommand{\lum}{erg~s\ensuremath{^{-1}}}

\newcommand{\msun}{\ensuremath{M_{\odot}}}
\newcommand{\kms}{\ensuremath{\mathrm{km~s^{-1}}}}
\newcommand{\mbh}{\ensuremath{M_\mathrm{BH}}}

\newcommand{\sigline}{ \ensuremath{\sigma_{\rm line} } }
\newcommand{\sgmint}{\ensuremath{\sigma_{\rm int}}}
\newcommand{\sgmtot}{\ensuremath{\sigma_{\rm tot}}}
\newcommand{\delmrm}{\ensuremath{\Delta M_{\rm BH}\rm(RM)}}
\newcommand{\mdelmrm}{\ensuremath{\langle\Delta M_{\rm BH}\rm(RM)\rangle}}

\newcommand{\chisq}{\ensuremath{\chi^2}}

\newcommand{\ha}{H\ensuremath{\alpha}}
\newcommand{\hb}{H\ensuremath{\beta}}

\newcommand{\oiii}{[O\,{\footnotesize III}]}

\newcommand{\feii}{Fe\,{\footnotesize II}}

\newcommand{\mgii}{Mg\,{\footnotesize II}}

\newcommand{\civ}{C\,{\footnotesize IV}}

\newcommand{\hst}{\emph{HST}}

\def\lax{{$\mathrel{\hbox{\rlap{\hbox{\lower4pt\hbox{$\sim$}}}\hbox{$<$}}}$}}
\def\gax{{$\mathrel{\hbox{\rlap{\hbox{\lower4pt\hbox{$\sim$}}}\hbox{$>$}}}$}}

\slugcomment{To appear in {\it The Astrophysical Journal}}
\shorttitle{Estimating \mbh\ with \mgii}
\shortauthors{Wang et al.}
\begin{document}

\title{Estimating Black Hole Masses in Active Galactic Nuclei Using the \mgii~$\lambda$2800 Emission Line}

\author{Jian-Guo~Wang\altaffilmark{1,2,5,6},
Xiao-Bo~Dong\altaffilmark{2,4},
Ting-Gui~Wang\altaffilmark{2,4},
Luis~C.~Ho\altaffilmark{3},
Weimin~Yuan\altaffilmark{1},
Huiyuan~Wang\altaffilmark{2,4},
Kai~Zhang\altaffilmark{2,4},
Shaohua~Zhang\altaffilmark{2,4}, and
Hongyan~Zhou\altaffilmark{2,4}
}

\altaffiltext{1}{National Astronomical Observatories/Yunnan
Observatory, Chinese Academy of Sciences, P.O. Box 110, Kunming,
Yunnan 650011, China;~wangjg,\,wmy@ynao.ac.cn}
\altaffiltext{2}{Key Laboratory for Research in
Galaxies and Cosmology, The University of Sciences and Technology of
China, Chinese Academy of Sciences, Hefei, Anhui 230026, China;
~xbdong,\,twang@ustc.edu.cn}
\altaffiltext{3}{The Observatories of the Carnegie Institution for Science,
813 Santa Barbara Street, Pasadena, CA 91101,
USA;~lho@obs.carnegiescience.edu}
\altaffiltext{4}{Center for Astrophysics, University of Science and Technology of China,
Hefei, Anhui 230026, China}
\altaffiltext{5}{Department of Physics, Yunnan University, Kunming,
Yunnan 650031, China}
\altaffiltext{6}{Graduate School of the
Chinese Academy of Sciences, 19A Yuquan Road, P.O. Box 3908, Beijing
100039, China}

\begin{abstract}

We investigate the relationship between the linewidths of broad \mgii\
$\lambda$2800 and \hb\ in active galactic nuclei (AGNs) to refine them as
tools to estimate black hole (BH) masses.  We perform a detailed spectral
analysis of a large sample of AGNs at intermediate redshifts selected from the
Sloan Digital Sky Survey, along with a smaller sample of archival ultraviolet
spectra for nearby sources monitored with reverberation mapping (RM).  Careful
attention is devoted to accurate spectral decomposition, especially in the
treatment of narrow-line blending and \feii\ contamination.  We show that,
contrary to popular belief, the velocity width of \mgii\ tends to be smaller
than that of \hb, suggesting that the two species are not cospatial in the
broad-line region.  Using these findings and recently updated BH mass
measurements from RM, we present a new calibration of
the empirical prescriptions for estimating virial BH masses for AGNs using the
broad \mgii\ and \hb\ lines.  We show that the BH masses derived from our
new formalisms show subtle but important differences compared to some of the
mass estimators currently used in the literature.
\end{abstract}


\section{Introduction}
It is generally accepted that active galactic nuclei (AGNs) are powered by the
release of gravitational energy from material accreted onto supermassive black
holes (BHs).  The determination of BH mass (\mbh) is crucial for understanding
the AGN phenomena, the cosmological evolution of BHs, and even the coevolution
of AGNs and their host galaxies.  Yet, for such distant objects, it is
currently impossible to obtain direct measurement of \mbh\ using spatially
resolved stellar or gas kinematics.  Fortunately, significant advances have
been made in recent years from reverberation mapping (RM) studies of nearby
Seyfert galaxies and quasi-stellar objects (QSOs; e.g., Wandel et al. 1999; Kaspi et al. 2000;
Peterson et al. 2004).  First, the anti-correlation between the radius of the
broad-line region (BLR) and the velocity width of broad emission lines for single objects supports
the idea that the BLR gas is virialized and that its velocity field is
dominated by the gravity of the BH (Peterson \& Wandel 1999, 2000; Onken \&
Peterson 2002).  Second, the size of the BLR scales with the continuum
luminosity (Kaspi et al. 2000, 2005), approximately as $R \propto L^{0.5}$
(Bentz et al. 2006, 2009); the $R-L$ relation offers a highly efficient
procedure for estimating the BLR size without carrying out time-consuming RM
observations.  And third, the BH masses estimated by RM are roughly consistent
(Gebhardt et al. 2000b; Ferrarese et al. 2001; Nelson et al. 2004; Onken et al.
2004) with the predictions from the tight correlation between \mbh\ and bulge
stellar velocity dispersion established for inactive galaxies (the \mbh--$\sigma_\star$ relation;
Gebhardt et al.  2000a; Ferrarese \& Merritt 2000).  These
developments imply that we can estimate the BH mass in type 1 (broad-line,
unobscured) AGNs by simple application of the virial theorem, \mbh\ = $fRv^2/G$,
where $f$ is a geometric factor of order unity that depends on the geometry and
kinematics of the line-emitting region, $R$ is the radius of the BLR derived
from the AGN luminosity, and $v$ is some measure of the virial velocity of
the gas measured from single-epoch spectra. The feasibility of obtaining
$R$ and $v$ from single-epoch spectra enables \mbh\ to be estimated very
efficiently for large samples of AGNs, especially for luminous quasars at
higher redshift that typically exhibit only slow and small-amplitude
variability (e.g., Kaspi et al. 2007), with the assumption that the virial relation
is independent of redshift and can be extrapolated to higher luminosities and masses.
In practice, for those AGNs that have
measurements of $\sigma_\star$, $f$ is determined empirically by scaling the
virial masses to the \mbh--$\sigma_\star$ relation of inactive galaxies
(e.g., Onken et al. 2004).  Implicit in this practice is the assumption---one
open to debate (Greene \& Ho 2006; Ho et al. 2008; Kim et al. 2008)---that
active and inactive BHs should follow the same \mbh--$\sigma_\star$ relation.
The most widely used estimator for $v$ is the full width at half-maximum (FWHM)
of the line.

Now, a large number of formalisms to estimate \mbh\ from single-epoch spectra
have been proposed in the recent literature, using different broad emission
lines optimized for different redshift regimes probed by (widely available)
optical spectroscopy.  At low redshifts, the lines of choice are \hb\ (Kaspi et
al. 2000; Collin et al. 2006; Vestergaard \& Peterson 2006) or \ha\ (Greene \&
Ho 2005). At intermediate redshifts, \mgii\ $\lambda$2800 is used (McLure \&
Jarvis 2002), while at high redshifts, one has to resort to \civ\ $\lambda$1549
(Vestergaard 2002; Vestergaard \& Peterson 2006). These formalisms are ultimately
calibrated against RM masses based on the \hb\ BLR radius and linewidth
measured from the variable (rms) spectra (Peterson et al. 2004).  Because the
\hb\ linewidth is typically smaller in the rms spectra than in the single-epoch
or mean spectra (Vestergaard 2002; Collin et al. 2006; Sulentic et al. 2006),
some authors have proposed that the FWHM used in the \hb-based formalisms
should be further corrected to obtain unbiased \mbh\ estimates (Collin et al.
2006; Sulentic et al. 2006).

As for the \mgii-based formalisms, because there are very few RM experiments
of the \mgii\ line, they are either based on the RM data for \hb\ (e.g., McLure
\& Jarvis 2002; McLure \& Dunlop 2004) or calibrated against the \hb\
formalisms themselves (e.g., Kollmeier et al. 2006; Salviander et al. 2007).
A strong, underlying assumption is that \mgii\ and \hb\ are emitted from the
same location in the BLR and have the same linewidth (see also Onken \&
Kollmeier 2008). In support of this assumption, some authors find that \mgii\
and \hb\ indeed have very similar linewidths (e.g., McLure \& Jarvis 2002;
McLure \& Dunlop 2004; Shen et al. 2008; also cf. Salviander et al. 2007).  However, there are
conflicting results in the literature: Corbett et al. (2003) claimed that \mgii\
is generally broader than \hb, whereas Dietrich \& Hamann (2004) came to an
opposite conclusion.  Certainly, the most direct way to settle this issue is
through direct RM of the \mgii\ line.  So far there are only two objects that
have successful \mgii\ RM, NGC\,5548 (Clavel et al. 1991; Dietrich \&
Kollatschny 1995) and NGC\,4151 (Metzroth et al. 2006).  These studies
tentatively suggest that \mgii\ responds more slowly to continuum variations
than \hb, implying that the \mgii-emitting region is larger than that radiating
\hb.

Thus, there are still some important open questions regarding the robustness
of \mbh\ measurements based on \mgii.  What is the relation between the
linewidths of \hb\ and \mgii?   Are estimates of \mbh\ based on \mgii\
consistent with those based on \hb?  These basic questions are critical for
understanding the systematic uncertainties in studies of the cosmological
evolution of BHs (cf. Shen et al. 2008; McGill et al. 2008; Denney et al. 2009a).
To address the above questions, we perform a detailed comparison of the
widths of the \mgii\ and \hb\ lines using single-epoch spectra for a large,
homogeneous sample of Seyfert 1 nuclei and QSOs at intermediate redshifts
culled from the Sloan Digital Sky Survey (SDSS; York et al. 2000).
We further compare single-epoch \mgii\ linewidths with \hb\ linewidths
measured from the rms spectra of AGNs with RM observations, finding systematic
deviations between the two.  We present a recalibration of the \mgii\ virial
mass estimator and compare our formalism with previous ones in the
literature.

This paper adopts the following set of cosmological parameters: $H_{\rm 0}$=70
km\,s$^{-1}$\,Mpc$^{-1}$, $\Omega_m $=0.3, and $\Omega_{\rm \Lambda}$=0.7.

\section{Sample and Data Analysis}

\subsection{The Samples}

The sample of best-studied \hb\ emission lines is the one compiled by Peterson
et al. (2004) for RM studies of 35 low-redshift AGNs. To compare \hb\ and \mgii\
for this sample, we located usable ultraviolet (UV) spectra for 29 sources, 16
from the {\it Hubble Space Telescope (HST)}\ and 13 from the {\it International
Ultraviolet Explorer (IUE)}\ data archives.  This sample will be used to study
the relationship between single-epoch \mgii\ linewidths and \hb\ linewidths
measured from rms spectra, and to fit a new \mbh\ formalism based on
single-epoch \mgii.

We also selected Seyfert 1 galaxies and QSOs in the redshift range $0.45<z<0.75$
from the Fifth Data Release (DR5) of the SDSS spectroscopic database
(Adelman-McCarthy et al. 2007).  Within this redshift range, both \hb\ and
\mgii\ lie within the SDSS spectral coverage.  To ensure accurate measurement
of both lines, we only select objects with a mean signal-to-noise ratio (S/N)
$\geq 20$ per pixel in both the \hb\ (4600--5100 \AA) and the \mgii\
(2700--2900 \AA) regions.
We discarded 26 spectra that have either broad absorption lines
or too many narrow absorption lines, or for which
 the \hb\ or the \mgii\ regions were corrupted by bad pixels.
The remaining 495 objects have spectra that can be well fitted, as
confirmed by visual inspection.
This sample will be used to investigate the FWHM relation between \mgii\ and \hb\
in single-epoch spectra and to compare our \mgii\ formalism with others in the literature.

\subsection{Spectral Fitting}

The spectra are first corrected for Galactic extinction using the
extinction map of Schlegel et al. (1998) and the reddening curve of
Fitzpatrick (1999).
Then the spectra are fitted using
an IDL code based on MPFIT (Markwardt 2009), which
performs $\chi^2$-minimization by the Levenberg--Marquardt technique.
Bad data are masked during the fitting.

To measure the \hb\ line, we perform continuum subtraction and emission-line
fitting following the method described in detail in Dong et al. (2008). We
first fit simultaneously the featureless nonstellar continuum (assumed to be a
power law), the \feii\ multiplet emission, and other emission lines in the
wavelength range 4200--5600 \AA, giving emphasis on the proper
determination of the local pseudocontinuum (continuum\,+\,\feii\
emission). For spectra with fits having a reduced $\chisq >1.1$
around \hb\ (4750--5050 \AA), a refined fit of the emission-line
profiles is performed to the pseudocontinuum-subtracted spectra
using the code described in Dong et al. (2005). Each
line of the \oiii\ $\lambda\lambda4959,5007$ doublet is modeled with
two Gaussians, one accounting for the line core and the other for
a possible blue wing as seen in many objects. The doublet lines
are assumed to have the same redshifts and profiles, and the flux
ratio $\lambda$5007/$\lambda$4959 is fixed to the theoretical value
of 3. The narrow component of \hb\ is fitted with one Gaussian, assumed to have
the same width as the line core of \oiii\ $\lambda5007$.  The broad component
of \hb\ is fitted with as many Gaussians as statistically
justified (see Dong et al. 2008 for details). \\

To measure the \mgii\ line, we adopt the following procedure.
We first obtain an initial estimate of the nonstellar featureless continuum by
fitting a simple power law,
\begin{equation}
f^{\rm PL}(\lambda;\, a, \beta) ~=~ a  \left(\frac{\lambda}{\rm 2200 \AA}\right)^{\beta} ~~,
\end{equation}
to the data in several continuum windows
near 2200, 3000, 4000, and 4200 \AA\ that suffer little from emission-line
contamination, if available. Then, the power-law local continuum, a Balmer continuum,
and an \feii\ emission template, which together constitute the so-called
pseudocontinuum, are fitted simultaneously.  The fitting is performed
in the restframe wavelength range 2200--3500 \AA, if available, with the small region
contaminated significantly by \mgii\ masked out.  The fitting range is set by
the wavelength coverage of the UV \feii\ template,
$f^{\rm T06}(\lambda)$,
which was generated by Tsuzuki et al. (2006) based on
their measurements of I\,Zw\,1.
In the wavelength region covered by \mgii\ emission,
they employed a semi-empirical iteration procedure to build the template.
They first generated a theoretical \feii\ model spectrum with
the photoionization code CLOUDY (Ferland et al. 1998) and subtracted it
from the observed I\,Zw\,1 spectrum around \mgii. Then
the \mgii\ doublet was fit assuming each line has the same profile
as \ha. And finally they obtained the \feii\ template underneath \mgii\
by subtracting the \mgii\ fit from the observed spectrum.
To match the linewidth and possible velocity shift of \feii\ lines,
we build the \feii\ model by convolving the I\,ZW\,1 template
with a Gaussian of width $\sigma_{\rm g}$ and shifting it with a velocity $v_{\rm shift}$
in logarithmic wavelength space (i.e., the velocity space because
$\mathrm{d}\ln\lambda = \mathrm{d}\lambda /\lambda = v/{\rm c}$),
as follows,
\begin{equation}
f^{\rm Fe\,II }(\lambda;\, c, v_{\rm shift}, \sigma_{\rm g} ) ~=~
c \, f^{\rm T06}(\lambda, v_{\rm shift}) \, \otimes \,
G(\lambda, \sigma_{\rm g}) ~~.
\end{equation}

As in Dietrich et al. (2002), the Balmer continuum is assumed to be produced in
partially optically thick clouds with a uniform temperature,
\footnote{We do not account for the velocity broadening of the Balmer
continuum, because the Balmer continuum in our fitting range is insensitive to
this effect.}
\begin{eqnarray}\label{balc_mod}
 f^{\rm BaC}(\lambda;\, d, T_e, \tau _{ \scriptscriptstyle \lambda}) & = &
 d ~ B_{\lambda}(\lambda,T_e)
   \left(1 - e^{-\tau _{ \scriptscriptstyle \lambda}  }\right);
   \quad \lambda \leq \lambda_{\rm BE}  \\
 \tau _{ \scriptscriptstyle \lambda} & = & \tau _{\rm  \scriptscriptstyle BE} ~
    \left({\lambda \over \lambda_{\rm BE}}\right)^{3},
\end{eqnarray}
\noindent
where $\lambda_{\rm BE} = 3646$ \AA\ (3.4 eV),
$\tau _{\rm  \scriptscriptstyle BE}$ is the optical depth at $\lambda_{\rm BE}$,
and $B_{\lambda}(\lambda,T_e)$ is the Planck blackbody spectrum at the
electron temperature $T_e$.

To sum up, the full model for the pseudocontinuum is as follows:
\begin{equation}
 f(\lambda) ~=~ f^{\rm PL}(\lambda; a, \beta) +
 f^{\rm Fe\,II}(\lambda; c, v_{\rm shift}, \sigma_{\rm g}) +
 f^{\rm BaC}(\lambda; d, T_e, \tau _{\scriptscriptstyle \rm BE}) ~~.
\end{equation}
The fitting is performed in logarithmic wavelength space.
During the fitting, the normalization $a$ and slope $\beta$
of the power-law continuum, the normalization $c$, velocity shift $v_{\rm shift}$
and broadening velocity $\sigma_{\rm g}$ of the \feii\ emission,
and the parameters $d$, $T_e$, and
$\tau _{\rm  \scriptscriptstyle BE}$ of the Balmer continuum
are set to be free parameters.

We note that in the fitting range of 2200--3500 \AA\ the Balmer continuum is
hard to be constrained and separated from the power-law continuum and \feii\
emission (cf. Figure 8 of Tsuzuki et al. 2006).  In this work, we are not
concerned with the properties of the Balmer continuum, but with the proper
separation of the power-law continuum, \feii, and \mgii.  To minimize the
effect of the possible poor fitting of Balmer continuum on
the determination of the power-law continuum, \feii, and \mgii,
we constrain the power-law continuum parameters in such a way that they vary only around
the best-fit values obtained from the first step, by a factor of $<10$\% for
the normalization $a$ and $<20$\% for the slope $\beta$. During this step, we
assign additionally larger weights to the regions 2400--2650 \AA\ and
2920--2990 \AA\ in order to improve the fit for the \feii\ emission surrounding
\mgii\ (cf. Section 2.3 of Dong et al. 2008).

Once the pseudocontinuum is fitted and subtracted, the \mgii\ emission line is fitted in
the range of 2700--2900 \AA, if available. For the SDSS sample, there are
a few cases where a small number of narrow absorption lines are present around the \mgii\
emission line. To further eliminate the absorption lines in these objects, we first fit
the \mgii\ emission line with one Gaussian, and then mask those pixels
of absorption features deviating strongly from
the model. The \mgii\ line is fitted in the following way. Each of the two \mgii\
$\lambda \lambda$2796,\,2803 doublet lines is modeled with two components, one
broad and the other narrow. The broad component is a truncated five-parameter
Gauss--Hermite series (van der Marel \& Franx 1993; see also Salviander et al.
2007); the narrow component is a single Gaussian. The broad components of the
doublet lines are set to have the same profile, with the flux ratio
$\lambda$2796/$\lambda$2803 set to be between 2:1 and 1:1 (Laor et al. 1997),
and the doublet separation set to the laboratory value.  The same prescription
is applied to the narrow components, with the following additional constraints:
FWHM $\leq 900$ \kms\ and flux $< 10\%$ of the total \mgii\ flux (Wills
et al. 1993; see also McLure \& Dunlop 2004).
The fitting results for all the 495 objects are reasonable
according to our visual inspection finally.
The FWHM value
is measured from the Gauss--Hermite model of \mgii\ $\lambda$2796.
The monochromatic flux of the continuum is measured from the fitted power law.

There are several other emission lines in the fitting region, identified from
the composite SDSS QSO spectrum (see Table 2 of Vanden Berk et al. 2001); yet,
because of their weakness, we simply masked them out in the fit.  Because of
the limited wavelength coverage of the RM sample, we cannot separate the Balmer
continuum from the power-law continuum. Thus, the Balmer continuum was not
included in the fits for this sample.  Additionally, there are deep narrow
absorption features around \mgii\ in the spectra of NGC~3227, NGC~3516,
NGC~3783, and NGC~4151 in the RM sample.  Each of the absorption features is
fitted simultaneously with a Gaussian when fitting the \mgii\ emission line.

We estimate the measurement uncertainties of the parameters using the bootstrap method%
\footnote{To estimate the errors on the fitted parameters, we generate 500
spectra by randomly combining the scaled model emission lines
of one object (denoted as ``A'') to the emission-line subtracted spectrum of
another object (denoted as ``B''). The emission-line model of object ``A''
is scaled in such a way that it has the same broad \mgii\ flux as object ``B,''
in order to minimize changes in S/N within the emission-line spectral regions
in the simulated spectra. Then, we fit the simulated spectra following
the same procedure as described in Section 2.2. For each parameter,
we consider the error typical of our sample to be the standard deviation
of the relative difference between the input and the recovered parameter values.
These relative differences turn out to be normally distributed
 for each of the parameters concerned.}
described in Dong et al. (2008, Section 2.5). The estimated 1 $\sigma$
errors for the broad-line fluxes are typically 10\% for \mgii\ and 8\% for \hb,
while the errors on the broad-line FWHM are $\sim$20\% for \mgii\ and $\sim$15\% for \hb.
The power-law continua have uncertainties of 8\% for the slope and 5\% for the
normalization. The above discussion does not account for possible systematic
errors resulting from the subtraction of the continuum or our treatment of the
\feii\ and narrow lines.

Figure 1 shows two examples of the fits.  The continuum and emission-line
parameters for the RM and SDSS samples are listed in Tables 1 and 2,
respectively.  The data and fitting parameters are available online for the
decomposed spectral components (continuum, \feii, and other emission lines).
\footnote{Available at \\
http://staff.ustc.edu.cn/\~{}xbdong/Data\_Release/MgII\_Hbeta/,
together with auxiliary code to explain the parameters and
to demonstrate the fitting.}

\subsection{Regression Methods}
In the next section, we will fit the linewidth relations and BH mass estimators
using several regression methods.
The purpose of using these different methods is for ease of comparison with
results in the literature, and to investigate possible differences in the
fitting results caused by the different methods.  Here we briefly summarize
the regression methods used.

\begin{enumerate}
\item{Ordinary least-squares (OLS), which is a least-squares regression method without considering measurement errors.}

\item{Weighted least-squares (WLS), which takes into account only measurement
uncertainties in the dependent variable.}

\item{FITexy (Press et al. 1992), which numerically solves for the minimum orthogonal
$\chi^2$ using an interactive root-finding algorithm. It accounts
for measurement uncertainties in both coordinates, but does not
account for intrinsic scatter.}

\item{FITexy$\_$T02, the version of FITexy modified by Tremaine et al. (2002),
accounts for possible intrinsic scatter in the dependent variable by
adding in quadrature a constant to the error value so as to obtain a
reduced \chisq\ of 1.}

\item{Gaussfit (McArthur et al. 1994), which implements generalized
least-squares using the robust Householder Orthogonal
Transformations (Jefferys 1980, 1981). It can handle errors in both coordinates,
but does not account for intrinsic scatter.}

\item{The bivariate correlated errors and intrinsic scatter (BCES) regression
method (Akritas \& Bershady 1996), which accounts for measurement
errors on both coordinates in the fit using bivariate correlated
errors, and possible intrinsic scatter (but does not output any
quantification of this scatter). The results of the two symmetrical
versions, bisector and orthogonal, are used in this paper.}

\item{LINMIX$\_$ERR (Kelly 2007), which accounts for measurement errors,
nondetections, and intrinsic scatter by adopting a Bayesian approach to compute the posterior
probability distribution of parameters, given observed data.
We also consider the
multivariate extension, MLINMIX$\_$ERR.}
\end{enumerate}

As we find below, most of the above regression methods
give consistent results.
For the linewidth-linewidth relationships, since
there is no prior knowledge about which variable is independent
and which is dependent,
we adopt the results given by the BCES orthogonal method,
which treats both variables symmetrically.
For the BH mass scaling relations, we adopt formally
the results given by the LINMIX$\_$ERR method,
since it is argued to be among the most robust regression methods
with the possibility of reliable estimation of intrinsic dispersion
(Kelly 2007).

For some of the regression methods listed above no intrinsic scatter (\sgmint)
can be inferred.
We can give a rough yet simple estimate of \sgmint$^2$\ by
deducting the contribution of the measurement errors from the variance in the regression residuals (\sgmtot$^2$), by
using an approximate relation
$\sgmint^2=\sgmtot^2-\langle\sigma_{\rm m}\rangle^2$,
where $\langle\sigma_{\rm m}\rangle$ is the median of the total measurement errors
computed from $\sigma_{\rm m}=\sigma_y+$slope$*\sigma_x$.
If \sgmtot\ is smaller than $\langle\sigma_{\rm m}\rangle$, \sgmint\ was set to be 0.
As a check, this rough estimate can be compared with the intrinsic scatter given
by some of the regression methods that provide such a measure.

\section{Results}

The main motivation of this work is to investigate whether reliable BH masses
can be estimated using the \mgii\ linewidth as a virial velocity indicator.
Linewidths are commonly parameterized as FWHM,
or sometimes as $\sigma_{\rm line}$---the line dispersion or
second moment of the line profile (Peterson et al. 2004).
Both quantities have
intrinsic strengths and weaknesses (see Section 3 of Peterson et al. 2004).
Collin et al. (2006), in particular, argued that the use of FWHM rather than
$\sigma_{\rm line}$ introduces systematic bias in \mbh\ estimates.
Although $\sigma_{\rm line}$ is a better tracer of virial velocity
than FWHM in rms spectra (Peterson et al. 2004),
the line dispersion is very sensitive to measurement errors
in the line wings, making it especially susceptible to inaccuracies caused by
deblending and subtraction of \feii\ and other emission lines, effects that
are particularly significant in mean and single-epoch spectra.  By contrast,
the FWHM is less prone to these effects; it is more sensitive to corrections
for the narrow-line component, which, fortunately, is quite weak for
\mgii\ (see Section 4.1).  In this work, we opt to use the FWHM to parameterize
the linewidth.

\subsection{Single-epoch \mgii\ FWHM versus \hb\ FWHM}

We first investigate the relation between the FWHM of \mgii\ and \hb, using
single-epoch data from our SDSS sample.  The relation is illustrated in Figure 2.
A strong correlation is present, but apparently deviates from one-to-one.
This trend has been noticed in the literature, but it was less prominent
because of the narrower dynamical range in velocity covered in previous studies
(e.g., Salviander et al. 2007; Hu et al. 2008).  With our high-quality data, we
can now fit a strict relation.  We perform a linear regression in log--log
space using the methods described in Section 2.3; the results are listed in
Table 3. As can be seen, most of the methods give mutually consistent results.
For our subsequent analysis, we adopt the BCES (orthogonal) method
because it treats both variables symmetrically (Section 2.3).  We find

\begin{eqnarray}
\nonumber
 \log \left[\frac{\rm FWHM(\mgii)}{1000~\kms}\right] ~&=&~ (0.81 \pm 0.02)
           \log \left[\frac{\rm FWHM(\hb)}{1000~\kms}\right] \\
 & &~ + ~ (0.05 \pm 0.01). ~~~~~
\end{eqnarray}

\noindent
This means that \emph{the line-emitting locations of \hb\ and \mgii\ in the
BLR are not identical}.
If they were, we would expect a linear relation between the two,
with no offset. The intrinsic scatter of this relation, as given by the
regression methods listed above,
is extremely small and negligible compared to the measurement errors.
The latter is actually comparable to the {\em total} scatter (\sgmtot)
of the relationship, which is found to be 0.08 dex.

\subsection{Single-epoch \mgii\ FWHM versus rms \hb\  $\sigma_{\rm line}$}

Since the assumption that \mgii\ FWHM is identical to \hb\ FWHM does not hold,
we explore the relation between \mgii\ FWHM and rms \hb\ $\sigma_{\rm line}$,
which has been argued to be a good tracer of the virial velocity of the BLR
clouds emitting (variable) \hb\ (see references in Section 1).  We use data for
the 29 objects in the RM sample that have UV spectra to perform this
exploration.  \mgii\ FWHM is measured from the single-epoch {\it HST/IUE}\
spectra, as listed in Table 1.  The data for  rms \hb\ $\sigma_{\rm line}$ are
mainly taken from Peterson et al. (2004).  In addition, we use updated RM data
for NGC\,4051 (Denney et al. 2009b), NGC\,4151 (Metzroth et al. 2006),
NGC\,4593 (Denney et al. 2006), NGC\,5548 (Bentz et al. 2007), and
PG\,2130+099 (Grier et al. 2008).  For objects with multiple measurements, the
geometric mean (i.e., the mean in the log scale) was used.

We find that the slope of the relation between \mgii\ FWHM and rms \hb\
$\sigma_{\rm line}$ deviates from unity, with a best fit of

\begin{eqnarray}
\nonumber
\log \left[\frac{\sigline({\rm \hb, rms})}{1000~\kms}\right] ~&=&~ (0.85\pm 0.21)
                         \log \left[\frac{\rm FWHM(\mgii)}{1000~\kms}\right] \\
 & & ~- ~ (0.21\pm0.12) ~~.
\end{eqnarray}

\noindent
The formal relation is nonlinear although the significance level is only
about 1\,$\sigma$.
A nonlinear relation between \mgii\ FWHM and rms \hb\ $\sigma_{\rm line}$
is not very surprising, in light of a similar situation observed for \hb\ FWHM
(Collin et al. 2006; also Sulentic et al. 2006, Section 1).
 For verification, we also fit the relation between \hb\ FWHM in the mean spectra
 and rms \hb\ $\sigma_{\rm line}$
 using data for 35 objects in the RM sample; the FWHM data
are taken from Collin et al. (2006) and from the updated sources
mentioned above.  The best-fit relation deviates from unity even
more seriously than the case of \mgii\ FWHM:

\begin{eqnarray}
\nonumber
\log \left[\frac{\sigline({\rm \hb}, {\rm rms})}{1000~\kms}\right] ~ &=&~ (0.54\pm 0.08)
                     \log \left[\frac{\rm FWHM(\hb,mean)}{1000~\kms}\right] \\
 & & ~- ~ (0.09\pm0.05) ~~.
\end{eqnarray}

\noindent
These relations between rms \hb\ $\sigma_{\rm line}$
and \mgii\ and \hb\ FWHM are illustrated in Figure 3.
The total 1 $\sigma$ scatter around these relationships
is \sgmtot\ = 0.12 dex for Equation (7) and 0.09 dex for Equation (8).
Given the relatively small measurement errors of the linewidths, there
likely exists intrinsic scatter in these relationships.
Using the simple method of deducting the measurement errors from
the total scatter, as described in Section 2.3,
we find \sgmint $\approx$ 0.09 dex and $\approx$ 0.08 dex for the relationships of
 Equations\,(7) and (8), respectively. The underlying reason for
the nonlinearity of these relationships may be, at least partially, that
the \sigline\ of rms spectra traces the velocity of the line-emitting region that
responds to continuum variation, while the FWHM of single-epoch spectra
may be contributed by various components (see Section 4.2 for a discussion).

\subsection{Practical Formalism for New \mgii-based \mbh\ Estimator}

As described above, \mgii\ FWHM is not identical to, but rather generally
smaller than, \hb\ FWHM;  for \mgii\ FWHM $\gtrsim 6000$ \kms, the difference
is $\gtrsim 0.2$\,dex.  This means that one of the fundamental
premises of the previous \mgii-based formalisms---that \mgii\ and \hb\
trace similar kinematics---does not hold.  Moreover, similar to the behavior
of \hb\ FWHM, \mgii\ FWHM seems not to be linearly proportional to rms \hb\
\sigline.  If rms \sigline\ is more directly linked to the virial velocity, this
implies that we cannot build a virial \mbh\ formalism by simply assuming
$\mbh \propto {\rm FWHM}^2$.  Furthermore, the \mbh\ data of the RM AGNs used
in McLure \& Jarvis (2002) and McLure \& Dunlop (2004) have since been
recalibrated or updated (Peterson et al. 2004; Denney et al. 2006, 2009b;
Metzroth et al. 2006; Bentz et al. 2007; Grier et al. 2008).  Thus, it is
necessary to reformulate the virial \mbh\ formalism based on single-epoch
\mgii\ FWHM.

We proceed by assuming that there is a tight relation between the BLR radius
of the \mgii-emitting region and the AGN continuum luminosity, in the form
$R_{\rm \mgii} \propto L^\beta$, and another between
the virial velocity of \mgii\ and the FWHM of the line,
in the form $v_{\rm virial}^2 \propto {\rm FWHM}^\gamma$.
Then, using the 29 objects with the \mbh\ values based on RM
and the \mgii\ data measured here (Table 1), we
calculate the free parameters by fitting

\begin{eqnarray}
\nonumber
\log \left[\frac{M_{\rm BH} {\rm (RM)}}{10^6 \, \msun}\right] ~&=&~  a + \beta
\log\left(\frac{L_{3000}}{10^{44}~\rm \lum}\right) \\
& &~ + ~ \gamma \log \left[\frac{\rm FWHM(\mgii)}{1000~\kms}\right] ~~,
\end{eqnarray}

\noindent
where $L_{3000} \equiv \lambda L_{\lambda}$(3000 \AA).  The RM-based \mbh\
data are mainly taken from Peterson et al. (2004), who calibrated the
$f$-factor by normalizing to the \mbh--$\sigma_\star$ relation of Onken et al.
(2004); \mbh\ for the updated objects comes from the references given in
Section 3.2.

We fit Equation (9) following four schemes, using the
(LINMIX$\_$ERR/MLINMIX$\_$ERR) method of Kelly (2007):
\begin{enumerate}
{\item $a$, $\beta$, and $\gamma$ are treated as free parameters.}

{\item $a$ and $\beta$ are treated as free parameters, but, as in all
previous formalisms, we fix $\gamma = 2$.}

{\item $a$ and $\beta$ are treated as free parameters, but we set
$\gamma = 1.70$, as suggested by  Equation (7).}

{\item $a$ and $\gamma$ are treated as free parameters, but we fix $\beta
= 0.5$, as suggested by the latest $R-L$ relation (Bentz et al. 2006, 2009).}
\end{enumerate}

Table 4 lists the best-fit regression for each scheme (Columns 1--4), as well as
comparisons between the \mbh\ estimates based on each scheme
and the RM-based masses
(Column 5).  It is apparent that the best-fit values for $\beta$ for all the
schemes are consistent with 0.5 within $1\,\sigma$ error.  Interestingly,
$\gamma$ appears to be marginally smaller than 2, since the standard deviation
of the BH mass for Scheme 2 is slightly larger than that for the other three
schemes.  If we set $\beta = 0.5$ (i.e., adopt Scheme 4), the best-fit
\mgii-based formalism is

%Eqn. 8
\begin{eqnarray}
\nonumber
\log \left(\frac{M_{\rm BH}}{10^6 \, \msun}\right) & = &  (1.13 \pm 0.27) +
0.5 \log\left(\frac{L_{3000}}{10^{44}~\rm \lum}\right)  \\
& & ~+~  (1.51 \pm 0.49) \log \left[\frac{\rm FWHM(\mgii)}{1000~\kms}\right] ~~.
\end{eqnarray}

Fitting the \hb\ FWHM data for the 35 RM objects under the same assumptions
($\beta = 0.5$), the \hb-based formalism becomes

%Eqn. 9
\begin{eqnarray}
\nonumber
\log \left(\frac{M_{\rm BH}}{10^6 \, \msun}\right) & = &  (1.39 \pm 0.14) +
0.5 \log\left(\frac{L_{5100}}{10^{44}~\rm \lum}\right)  \\
& & ~+~  (1.09 \pm 0.23) \log \left[\frac{\rm FWHM(\hb)}{1000~\kms}\right] ~~.
\end{eqnarray}

\noindent
The best-fitting $\gamma = 1.09 \pm 0.23$ agrees well with the
${\rm \sigline(\hb,rms)-FWHM(\hb)}$ relation derived in Equation (8).

Comparisons between the RM-based masses and
the \mbh\ estimates from our new \mbh\ formalisms using \mgii\ FWHM (Equation
(10)) and \hb\ FWHM (Equation (11)) are illustrated in Figure 4.
Following Vestergaard \& Peterson (2006), we calculate
the deviation of the new calibrated single-epoch \mbh\ estimates from the
RM-based masses, \delmrm.
The mean of the deviations,
\mdelmrm, is only 0.01 dex
for our \mgii\ estimator,
and the 1 $\sigma$ scatter is 0.4 dex (Column 5 in Table 4).
As a comparison, if we use the formalism of McLure \& Dunlop (2004),
the deviations from the same  RM-based masses
have a mean of 0.38 dex and a 1 $\sigma$ scatter of 0.45 dex.
For our \hb\ estimator,
 \mdelmrm\ is 0.01 dex and the 1 $\sigma$ scatter is 0.3 dex,
compared to \mdelmrm\ = 0.05 dex and $\sigma$ = 0.4 dex  if the
formalism of Vestergaard \& Peterson (2006) is used.%
\footnote{
It should be noted that here we use the averaged spectrum for an object
with more than one observation, unlike in Vestergaard \& Peterson (2006)
where the individual single-epoch \hb\ spectral data were used in the regression.
If we take the latter approach,
our \hb\ estimator gives a \mdelmrm\ = $-$0.07 dex
and a scatter of 0.33 dex,
while  Vestergaard \& Peterson (2006) gave \mdelmrm\ = $-$0.12 dex
and a scatter of 0.45 dex.}
It should be noted that these scatters of the scaling relations,
which give a measure of the uncertainty in estimating \mbh\ from
the single-epoch spectroscopic data, is relative to the RM-based masses only.
Since, as pointed out by Vestergaard \& Peterson (2006),
 the RM-based masses themselves are  uncertain typically by a factor of
$\sim 2.9$ (as calibrated against the $\mbh$--$\sigma_\star$ relation; Onken et al. 2004),
the absolute uncertainty of the masses thus estimated is even higher.
For the \hb\ formalism, we find this absolute uncertainty to be a factor of $\sim 3.5$, to be compared with a factor of
$\sim 4$ given in Vestergaard \& Peterson (2006);
for \mgii, we estimate that the absolute uncertainty is a factor of $\sim 4$.

As shown above, our new formalisms improve somewhat the scatter in the single-epoch
\mbh\ estimates compared to previous \hb\ and \mgii\ estimators, by 0.1 dex and 0.05 dex,
respectively. Given the same linewidth and luminosity data used in this work
and in  Vestergaard \& Peterson (2006), the reduction in the scatter of the \mbh\ estimates
should result from a decrease in the {\em intrinsic} dispersion of our improved single-epoch
\mbh\ formalisms. Using the LINMIX$\_$ERR method, the intrinsic scatter inherent in our \mbh\
formalism can be inferred to be 0.08 dex (1 $\sigma$) for the \hb\ and 0.14 dex for \mgii.

Figure 5 compares our new \mgii-based formalism (we show only Schemes 2 and 4)
with the previous \hb-based formalisms of Vestergaard \& Peterson (2006;
panel a) and Collin et al. (2006; panel b), as well as our newly
derived version using the SDSS sample of 495 Seyfert 1s and QSOs (Equation (11);
 panel c).  The \mbh\ residuals between our \mgii\ formalism
and the \hb\ formalisms are listed in Table 4 (Columns 6--8).
While our \mgii-based formalism, especially for Scheme 4 (Equation (10)), agrees
well with our \hb-based formalism (Equation (11)), note that it deviates markedly
from the \hb\ formalism of Vestergaard \& Peterson.  This
confirms previous suspicions (Collin et al. 2006; Sulentic et al. 2006) that
the use of \hb\ FWHM from mean and single-epoch spectra with the
assumption $\gamma = 2$ introduces systematic bias into \mbh\ estimates.

We further compare our new \mgii-based formalism (Equation (10)) with other \mgii\
formalisms widely used in the literature.
Figure 5 illustrates that the \mbh\ estimates following the formalisms of
McLure \& Dunlop (2004; panel d), Kollmeier et al. (2006; panel e),
and Salviander et al. (2007; panel f) show large systematic deviations,
mostly in the sense of being smaller than ours.
The deviations stem primarily from the recalibration of the RM masses;
other factors are discussed in Section~4.3.
We note that a yet-unpublished \mgii-based formalism by
M.~Vestergaard et al. (in preparation)
used in the recent literature (e.g., Kelly et al. 2009)
is almost identical to our Scheme 2 (with $\gamma$ fixed to 2).

\section{Discussion}

\subsection{Testing the Effect of Narrow-line Subtraction}

The narrow component of \mgii\ is generally weak in luminous type~1 AGNs
(e.g., Wills et al. 1993; Laor et al. 1994), and so its contribution to the
total line flux can be safely neglected.  However, its presence might have a
more pronounced impact on the FWHM measurement of broad \mgii.  In the
literature, narrow \mgii\ was accounted for in the line fitting by some
authors (e.g., McLure \& Dunlop 2004), but not by others (e.g., Salviander et
al. 2007).  As there is usually no clear inflection in the \mgii\ profile,
separating narrow \mgii\ from the broad component is often challenging.
Fortunately, for the spectra in our SDSS sample,
\oiii\ $\lambda$5007 is present, and thus we can use \oiii\ to try to
constrain narrow \mgii, to test the effect of narrow \mgii\ on
the FWHM measurement of broad \mgii, and also to test the reliability of
our \mgii\ fitting strategy.

In addition to the default fitting strategy described in Section 2.2,
in which narrow \mgii\ is modeled as a single free Gaussian, we
tried two alternative strategies in which narrow \mgii\ is (A) not
fit at all, and (B) is fit using a single-Gaussian model constrained
to that of the line core of \oiii.  Broad \mgii\ is modeled as
described in Section 2.2.  We find that, for the 495 objects in our SDSS sample,
the distributions of the
reduced \chisq\
of the \mgii\ emission line fit
of the three approaches can be approximated
reasonably well with a log-normal function.  The peak and standard
deviation of the reduced \chisq\ are very similar for all three,
being (0.97, 0.10 dex) for the default strategy, (1.01, 0.10 dex) for
Strategy A, and (0.99, 0.10 dex) for Strategy B.
Regarding the FWHM of broad \mgii,
the mean and standard deviation are
($-0.04$, 0.05) for $\log [\frac{\rm FWHM(A)}{\rm FWHM(default)}]$ and
(0.00, 0.05) for $\log [\frac{\rm FWHM(B)}{\rm FWHM(default)}]$.
For Strategy B, the fitted flux of narrow \mgii\ is
less than 10\% of the total line flux for almost all the objects.
Our tests show that omitting the subtraction of narrow \mgii\ has a
negligible effect on the FWHM of broad \mgii, typically decreasing
it only by a tiny factor of 0.04 dex.  We further confirm that our
default procedure for modeling narrow \mgii\ is consistent with that
using the \oiii\ core as a template.

The above Strategy A is exactly the same as the \mgii-fitting method adopted
by Salviander et al. (2007). We also compared our method with that of
McLure \& Dunlop (2004). When fitting the spectra in our SDSS sample by the
method of McLure \& Dunlop (2004), on average the FWHM of broad \mgii\ is
larger than that of our method by 0.1 dex.

\subsection{\mbh\ Estimators with Single-epoch \hb\ and \mgii}

As analyzed in detail by Sulentic et al. (2006), the overall profile of the
\hb\ emission line, as viewed in single-epoch spectra, likely comprises
multiple components emitted from different sites.  First, as a recombination
line, \hb\ can arise from BLR gas that is very close to the central engine.
Then \hb\ can be gravitationally redshifted, as (part of) the component of
the ``very BLR'' (Marziani \& Sulentic 1993).
Such clouds may be optically thin to the ionizing continuum, such that
\hb\ is no longer responsive to continuum variation (Shields et al. 1995).
Second, like \civ\ $\lambda$1549, \hb\ can be produced partly in
high-ionization winds, as some observations suggest (see Marziani et al.
2008, and references therein).
This wind component would not be virial.
Third, \hb\ can also be produced on the surface of the accretion disk,
both by recombination and collisional excitation (Chen \& Halpern 1989;
Wang et al. 2005; Wu et al. 2008); this component would be highly anisotropic
(cf. Collin et al. 2006).  Considering the above factors, it is not surprising
that single-epoch FWHM is not linearly proportional to \sigline\ for rms \hb\
(Equation (8)).

\mgii, as a low-ionization, collisionally excited emission line, cannot be
produced in clouds very close to the central engine.  Furthermore, because
\mgii\ originates only from optically thick clouds, radiation pressure force
cannot act on them very significantly (cf. Marconi et al. 2008, 2009; Dong et
al. 2009a,b), and thus \mgii\ suffers little from nonvirial motion.
Hence, compared to \hb, the FWHM of single-epoch \mgii\ should, in principle,
deviate less, if at all, from the true virial velocity of the line-emitting
clouds.  This is suggested by the best-fit value for $\gamma$ in Equation (10),
which indicates ${v_{\rm virial}^2} \propto {\rm FWHM(\mgii)}^{1.51\pm0.49}$.

Previously, researchers have feared that the substantial contamination
of the \mgii\ region by \feii\ multiplets might introduce significant
uncertainties in its linewidth measurements, such that \mgii-based \mbh\
estimates may not be as accurate as those based on \hb.  With the recent
availability of a more refined UV \feii\ template (Tsuzuki et al. 2006; cf.
Vestergaard \& Wilkes 2001), we have higher confidence that the linewidth
measurements of \mgii\ are reasonably robust.  Our work suggests that we can
measure \mgii\ FWHM typically to within an uncertainty of $\sim$20\%.
Nevertheless, it would be highly desirable to attempt to further improve
the methodology for \feii\ subtraction, not only in the UV but also at
optical wavelengths.

\subsection{Comparison with Previous Studies}
As shown in Section\,3 (see Figure 5), our \mgii- and \hb-based \mbh\ formalisms show,
in addition to somewhat improved internal scatter,
subtle but systematic deviations from some of the commonly
used \mbh\ estimators in the literature. In general, the formalism prescribed
by our Scheme 4 (Equation (10); $\mbh \propto {\rm FWHM}^{1.51 \pm 0.49}$) gives
progressively higher and lower \mbh\ values toward the low- and high-mass
ends, respectively.  The only exception is the \hb-based formalism of Collin
et al.  (2006), which gives roughly consistent results as ours over a
relatively large mass range.
The discrepancies between previous mass
estimators and ours arise from one, or a combination, of the following factors
incorporated into our analysis. (1) We use the most recently recalibrated and
updated RM \mbh\ measurements from
the literature (Peterson et al. 2004; Denney et al. 2006, 2009b;
 Metzroth et al. 2006; Bentz et al. 2007; Grier et al. 2008).
(2) Our new formalism (Scheme 4, Equation\,(10)) uses the
best-fitting value of $\gamma$ instead of the canonical value of $\gamma=2$.
(3) For \mgii, we determine the scaling factor (incorporated into the
coefficient $a$ of Equation\,(9)) and the power-law index ($\beta$) of the
$R_{\rm \mgii}$--L relation by fitting Equation\,(9) to the data, instead of simply
using the existing $R_{\rm \hb}$--L relation as a surrogate.
(4) Differences in the line-fitting and determination of the FWHM.
We discuss each of these factors in detail below.

Specifically,
assuming the canonical value of $\gamma=2$ in Equation\,(9)  would, compared to our
Scheme 4, underestimate \mbh\ at the low-end and overestimate \mbh\ at the
high-end,  for both \mgii\ and \hb.  This accounts for most of the deviations
from Vestergaard \& Peterson (2006) and Kollmeier et al. (2006), and partially
from others in Figure 5.
In order to account for systematic biases with
respect to RM-based \mbh, Collin et al. (2006) introduced a
correction factor, which is dependent on \hb\ FWHM, into their
\hb-based formalism assuming $\gamma=2$. This correction has a
similar effect as fitting $\gamma$ as a free parameter, as we do here,
and thus the rough consistency between our results and theirs is not
surprising. Factor (3) is also partially responsible for producing
the deviations from some of the previous \mgii-based \mbh\
estimators, such as those of McLure \& Dunlop (2004) and Salviander et al.
(2007), who assumed that both \mgii\ and \hb\ obey the same $R$--$L$ relation,
and of Kollmeier et al. (2006), who used a very steep relation of
$R_{\rm \mgii} \propto L^{0.88}$.

There have been previous reports of discrepancies between \mgii- and \hb-based
estimators, which are sometimes claimed to correlate with luminosity or
Eddington ratio (e.g., Kollmeier et al. 2006; Onken \& Kollmeier 2008). These
effects can be traced, at least partially, to the one-to-one relation assumed
between FWHM(\mgii) and FWHM(\hb), which is
contradictory to the nonlinear relation found in this work. In fact, by
adopting a nonlinear FWHM(\mgii)--FWHM(\hb) relation and $R_{\rm \mgii}
\propto L^{0.5}$, our new \mgii- and \hb-based estimators yield mutually
consistent results for the SDSS sample (Figure 5, panel c).   We verified
that the previously claimed correlations of the residuals of the \mgii-
and \hb-based estimators with luminosity or Eddington ratio largely vanish; a
Spearman rank analysis indicates a chance probability of 0.09 for the former
and 0.05 for the latter.

It is generally accepted that the width of the variable part of the line,
the line dispersion $\sigma_{\rm line}$ measured in the rms spectrum, is by
far the best tracer of the virial velocity of the BLR gas
responsible for
the variable portion of the emission line (e.g., Onken \& Peterson 2002), such that
$\mbh \propto \sigma_{\rm line}^2$ according to the virial theorem.
If the virial velocity is estimated using
FWHM (or any other measure of the linewidth) in single-epoch spectra, as long
as its relation with $\sigma_{\rm line}$ is nonlinear, $\gamma$ in Equation\,(9) is
expected to deviate from $\gamma=2$. This is exactly what we find in this work
for FWHM(\mgii) at $1\,\sigma$ significance level,
as well as for FWHM(\hb) at $6\,\sigma$ significance level.
In fact, the fitted value of
$\gamma=1.51\pm0.49$ for \mgii\ (Equation\,(10)) and $1.09\pm0.23$ for  \hb\ (Equation\,(11))
are almost identical to those derived from the fitted
$\sigma_{\rm line}$--FWHM(\mgii) and $\sigma_{\rm line}$--FWHM(\hb) relations,
which have slopes of $1.70\pm0.42$ and $1.08\pm0.16$, respectively.
Factor (3) is justified by the compelling evidence presented in this work that
the line-emitting locations of \hb\ and \mgii\ in the BLR are not identical.
Possible physical processes underlying factors (2) and (3) are discussed in
Section 4.2.

As an additional consideration, we have performed in this work refined and careful line fitting
and determination of the FWHM, which may have subtle differences from
previous results. These differences may also give rise to, to some extent,
the systematic discrepancies between the mass relations in Figure 5
since we use our measured FWHM and luminosity data when producing the figure.
For example, Salviander et al. (2007) did not subtract the narrow component of \mgii,
leading to \mgii\ FWHM  statistically smaller than ours; so the true deviations in
Figure 5 (panel  f) would be larger if their FWHM data were used.
On the contrary, McLure \& Dunlop (2004) over-subtracted the narrow component
compared to ours (they considered a possible narrow component as
having an upper limit of FWHM = 2000 \kms, much larger than the 900 \kms\ used in our work),
and their \mgii\ FWHM are  statistically larger than ours; thus, the true deviations in Figure 5
(panel e) would be somewhat smaller if their FWHM  data were used.

Finally, the appropriateness of our approach is further justified
by the fact that our formalisms give \mbh\ values consistent with
the RM measurements with the least systematic bias,
as well as a reduced (intrinsic) scatter compared to previous formalisms (Section 3.3).
Moreover, we find consistent masses between the \mgii- and \hb-based estimators.  We thus
conclude that our \mbh\ estimators introduce less systematic bias
compared to previous formalisms. Obviously, more RM measurements (for both
\hb\ and \mgii) are needed in order to improve the determination of the
$\sigma_{\rm line}$--FWHM relation, the $R$--$L$ relation, and the index
$\gamma$ in the $\mbh \propto {\rm FWHM}^{\gamma}$ relation.

\section{Summary and Conclusions}

We investigate the relation between the velocity widths for the broad \mgii\
and \hb\ emission lines, derived from FWHM measurements of single-epoch spectra
from a homogeneous sample of 495 SDSS Seyfert 1s and QSOs at $0.45 < z <0.75$.
Careful attention is devoted to accurate spectral decomposition, especially
in the treatment of narrow-line blending and \feii\ contamination.
We find that \mgii\ FWHM is systematic smaller than \hb\ FWHM, such that
FWHM(\mgii)\,$\propto$\,FWHM(\hb)$^{0.81\pm 0.02}$.  Using 29 AGNs that have
optical RM data and usable archival UV spectra, we then
investigate the relation between single-epoch \mgii\ FWHM and rms \hb\
\sigline\ (line dispersion), a quantity regarded as a good tracer of the
virial velocity of the BLR clouds emitting the variable \hb\ component.
We find that, similar to the situation for the FWHM of single-epoch \hb,
single-epoch \mgii\ FWHM is unlikely to be linearly proportional to rms \hb\
\sigline.  The above two findings suggest that a major assumption of previous
\mgii-based virial BH mass formalisms---that the \mgii-emitting region is
identical to that of $\hb$---is problematic.  This finding and the recent
updates of the reverberation-mapped BH masses (Peterson et al. 2004; Denney et
al. 2006, 2009b; Metzroth et al. 2006; Bentz et al. 2007; Grier et al. 2008)
motivated us to recalibrate the \mbh\ estimator based on single-epoch \mgii\
spectra.

Starting with the empirically well-motivated BLR radius--luminosity relation
and the virial theorem, $\mbh \propto L^\beta {\rm FWHM}^\gamma$,
we fit the reverberation-mapped objects in a variety of different ways to
constrain $\beta$ and $\gamma$.  For all the strategies we have considered,
$\beta$ has a well-defined value of $\sim$0.5, in excellent agreement with
the latest BLR radius--luminosity relation (Bentz et al. 2006, 2009), whereas
$\gamma \approx 1.5 \pm 0.5$, which is marginally in conflict with the
canonical value of $\gamma = 2$ normally assumed in past studies.
Performing a similar exercise for \hb\ yields $\mbh \propto L^{0.5}
{\rm FWHM(\hb)}^{1.09\pm0.22}$, which again significantly departs from
the functional forms used in the literature.
The 1 $\sigma$ uncertainty (scatter) is of the order of 0.3 dex relative to the
RM-based masses for the \hb\ estimator, and $\sim$0.4 dex for the \mgii\ estimator.
Using the same data set, the scatter of our \hb\ mass scaling relation is reduced by  0.1 dex
over that of Vestergaard \& Peterson (2006), indicating improvement in the internal scatter.

We use the SDSS database to compare our new \mbh\ estimators with various
existing formalisms based on single-epoch \hb\ and \mgii\ spectra.  BH masses
derived from our \mgii-based mass estimator show subtle but important
deviations from many of the commonly used \mbh\ estimators in the
literature.  Most of the differences stem from the recent recalibration of the
masses derived from RM.  Researchers should exercise
caution in selecting the most up-to-date \mbh\ estimators, which are
presented here.

%%%%%%%%%%%%%%%%%%%%%%%%%%%%%%%%%%%%%%%%%%%%%%%%%%%%%%%%%%%%
\acknowledgments We thank the referee, Michael Strauss, for his
careful comments and helpful suggestions that improved the paper.
This work is supported by Chinese NSF grants NSF-10533050 and
NSF-10703006, National Basic Research Program of China (973 Program
2009CB824800) and a CAS Knowledge Innovation Program (grant no.
KJCX2-YW-T05). The research of L.C.H. is supported by the Carnegie
Institution for Science. Funding for the SDSS and SDSS-II has been
provided by the Alfred P. Sloan Foundation, the Participating
Institutions, the National Science Foundation, the U.S. Department
of Energy, the National Aeronautics and Space Administration, the
Japanese Monbukagakusho, the Max Planck Society, and the Higher
Education Funding Council for England. The SDSS Web Site is
http://www.sdss.org/.
%%%%%%%%%%%%%%%%%%%%%%%%%%%%%%%%%%%%%%%%%%%%%%%%%%%%%%%%%%%%

%\clearpage
%%%%%%%%%%%%%%%%%%%%%%%%%%%%%%%%%%%%%%%%% figures

\begin{figure}[tbp]
\label{fig-1}\epsscale{1} \plotone{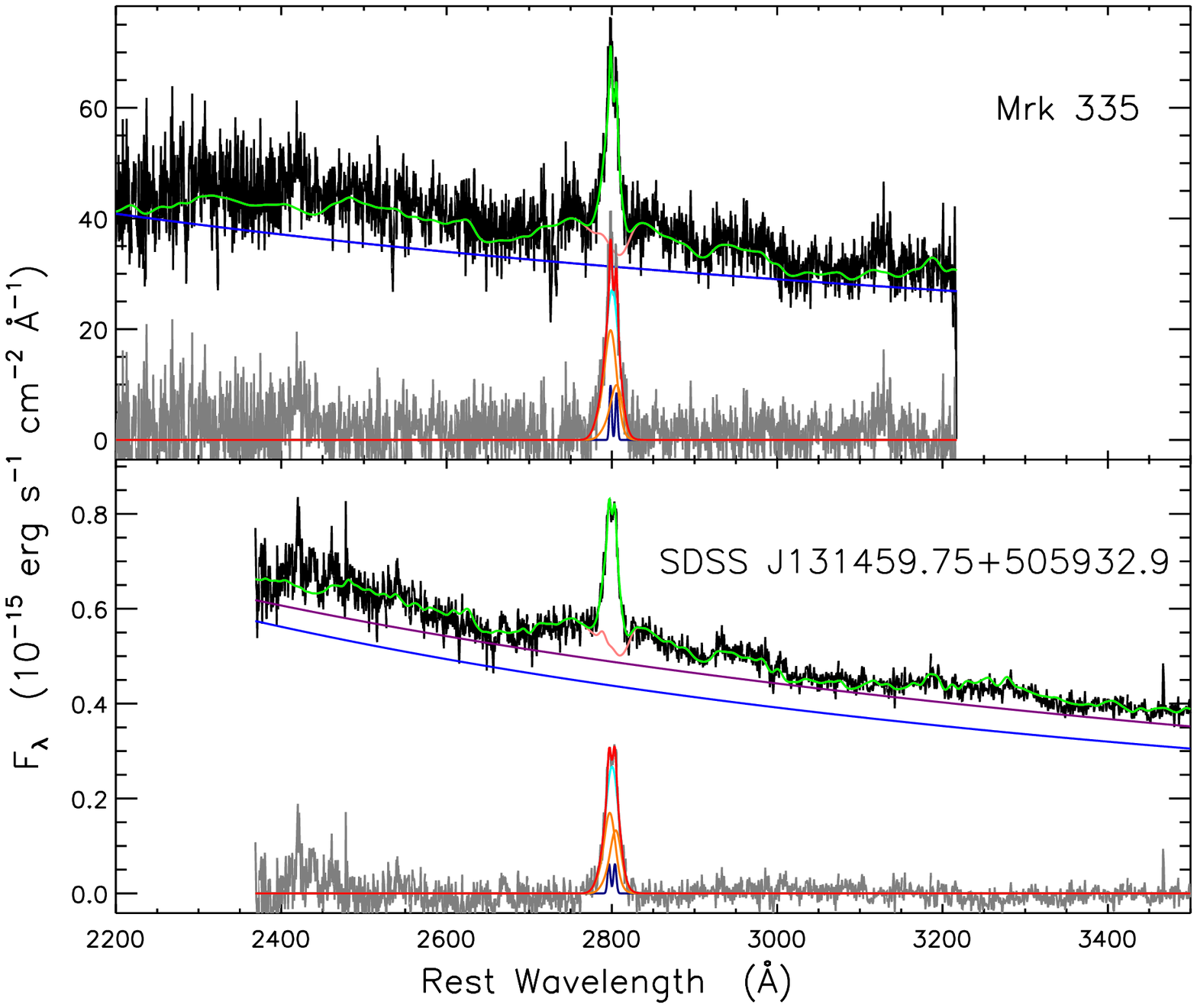} \caption{Examples of
\mgii\ fitting for ({\it top}) the \hst\ spectrum of Mrk 335 and
({\it bottom}) the SDSS spectrum of SDSS~J131459.75+505932.9.  The data are
shown in black, the power-law AGN continuum in blue, the pseudocontinuum
(power law plus \feii\ emission) in pink, the final model for all fitted
components in green, and the continuum-subtracted emission-line spectrum in
gray.  For the multi-Gaussian fit to \mgii, the
narrow components are shown in navy, the individual broad components in brown,
the sum of all the broad components in cyan, and the total model (narrow plus
broad) in red.}
\end{figure}

%\clearpage
%%%%%%%%%%%%%%%%%%%%%%%%%%%%%%%%%%%%%%%%% figures

\begin{figure}[tbp]
\label{fig-2}\epsscale{1} \plotone{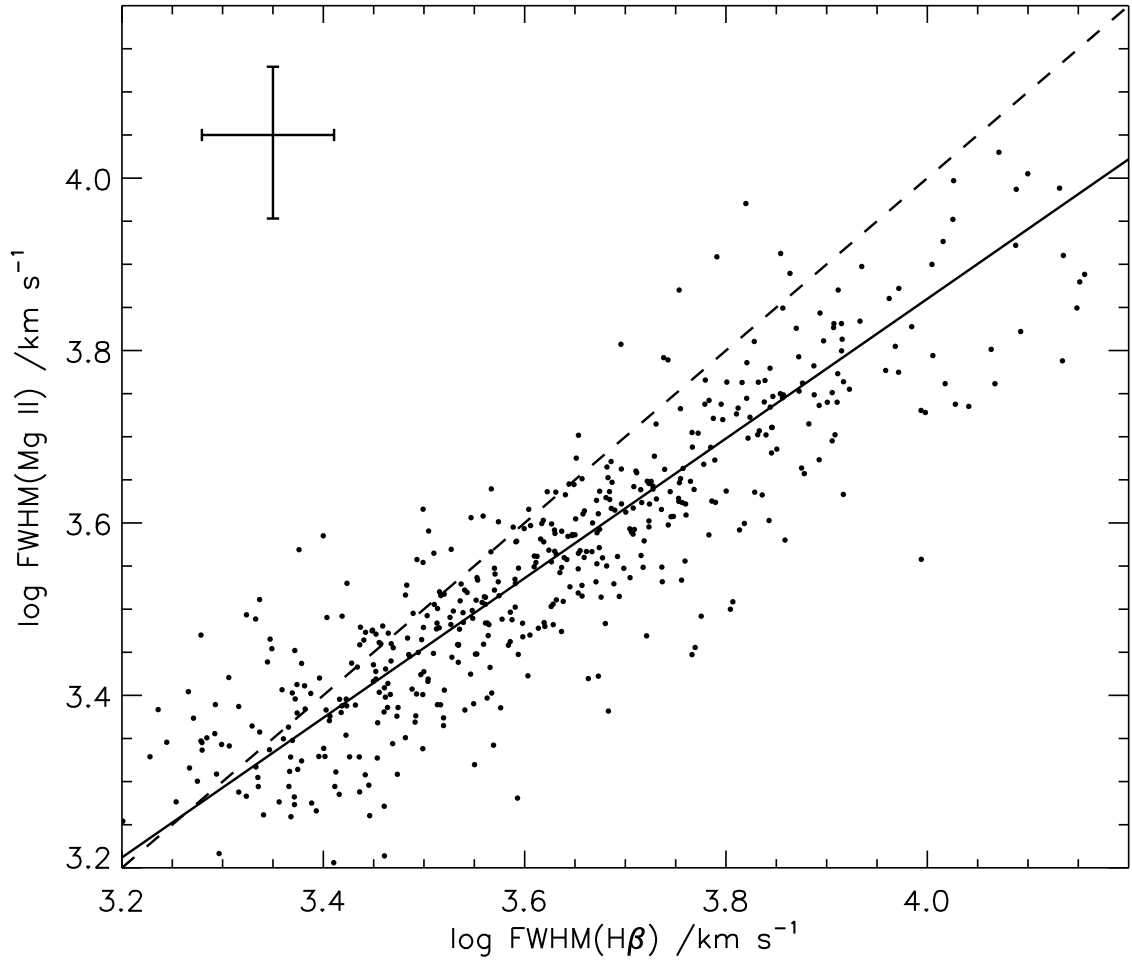} \caption{FWHM(\mgii)
vs. FWHM(\hb) for our SDSS sample. The solid line represents the
best-fitting power law with index 0.81. The dashed line represents a 1:1
relationship.
A typical 1 $\sigma$ error bar is also shown (top-left).
}
\end{figure}

%\clearpage
%%%%%%%%%%%%%%%%%%%%%%%%%%%%%%%%%%%%%%%%% figures

\begin{figure}[tbp]
\label{fig-3}\epsscale{1} \plotone{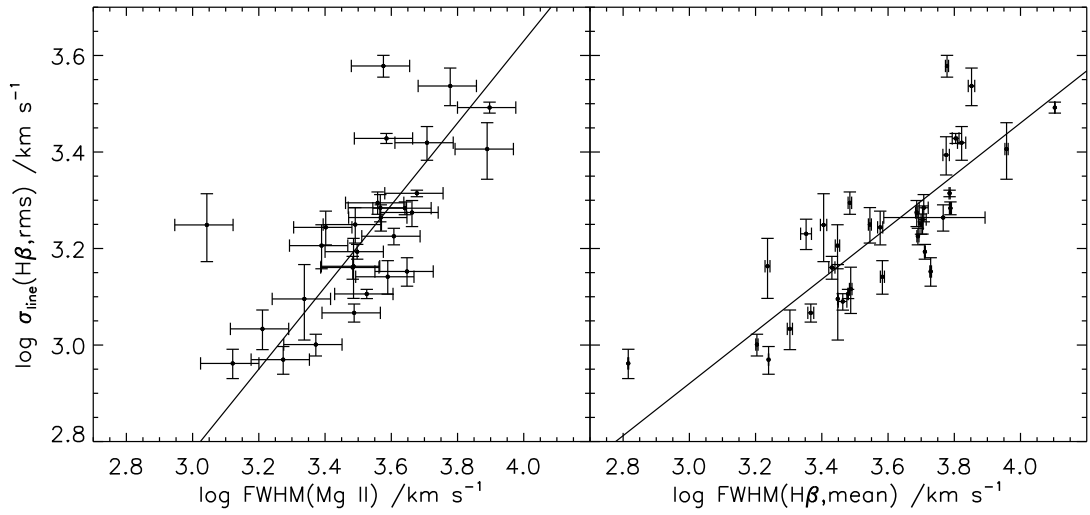}
\caption{ $\sigma_{\rm line}({\rm
\hb, rms})$ vs.  FWHM(\mgii)
of the 29 objects with \mgii\ FWHM measured in the paper (left panel)
and FWHM(\hb, mean)
of the 35 objects from Collin et al. (2006) and recent updated data  (right panel).
The solid lines show the best-fitting relations.
The error bars are at 1 $\sigma$.}
\end{figure}

%\clearpage
%%%%%%%%%%%%%%%%%%%%%%%%%%%%%%%%%%%%%%%%% figures

\begin{figure}[tbp]
\label{fig-4}\epsscale{1} \plotone{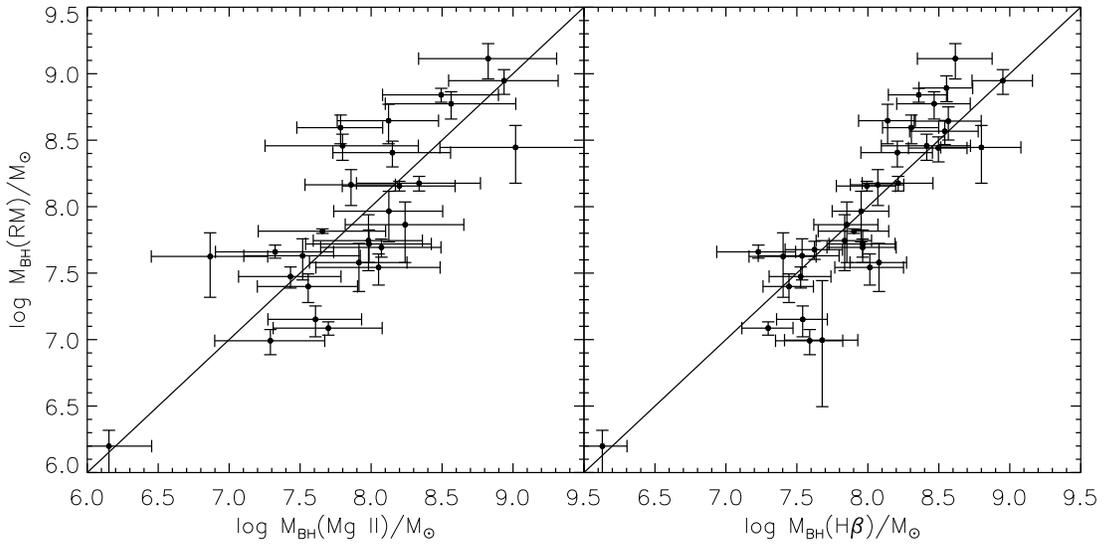}
\caption{BH masses estimated
from RM plotted against masses obtained from
\mgii\
(left panel; using the 29 objects with \mgii\ data measured in the paper),
and from \hb\
(right panel; using the 35 objects from Peterson et al. (2004) and recent updated data).
The solid line represents a 1:1 relationship.
The error bars are at 1 $\sigma$.}
\end{figure}

%\clearpage

%%%%%%%%%%%%%%%%%%%%%%%%%%%%%%%%%%%%%%%%% figures

\begin{figure}[tbp]
\label{fig-5}\epsscale{1} \plotone{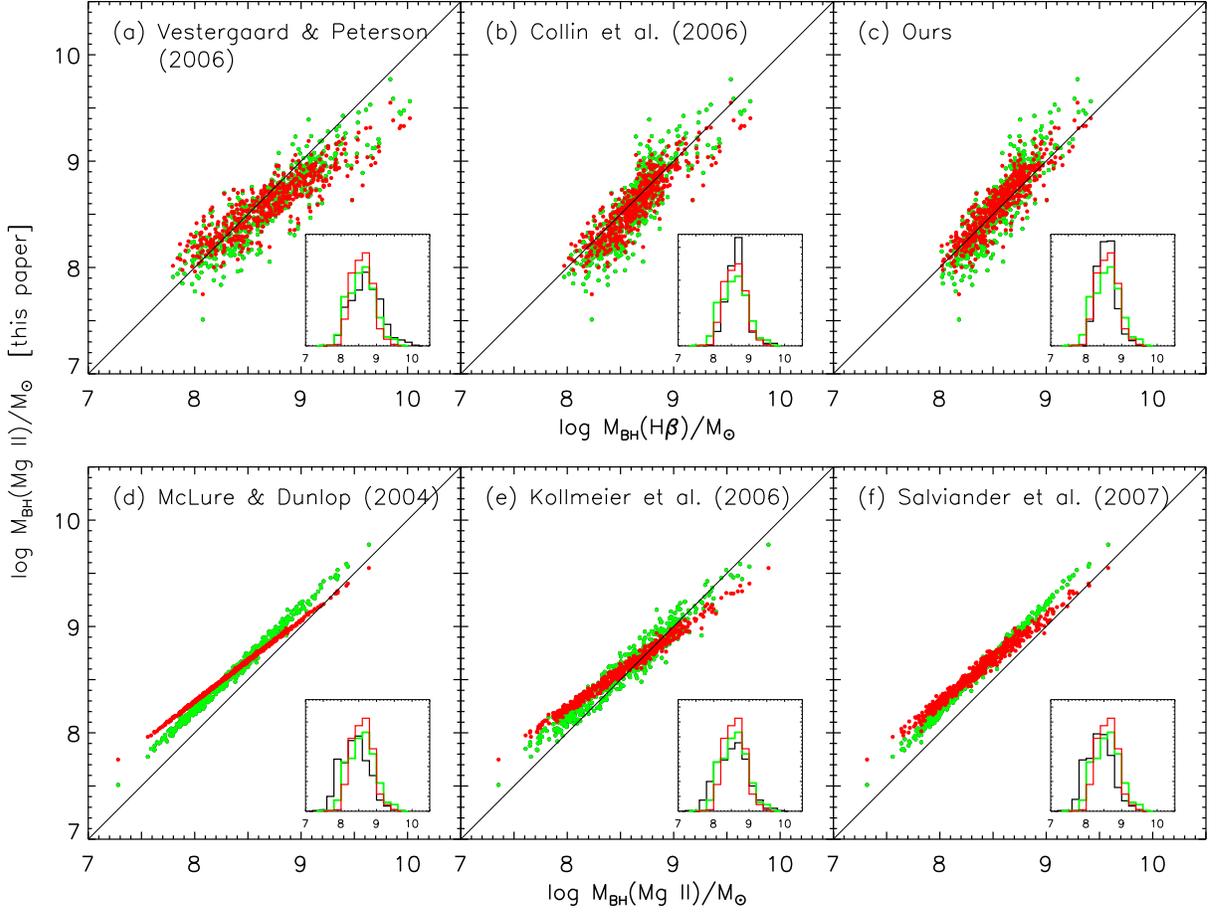} \caption{Comparison of \mbh\
estimated by our \mgii\ formalism with other formulae discussed in Section 3,
using the SDSS sample.
The $y$-coordinates of the green and red points represent \mgii\ masses
estimated by Scheme 2
($\mbh \propto {\rm FWHM}^{2}$) and Scheme 4
($\mbh \propto {\rm FWHM}^{1.51 \pm 0.49}$), respectively.
The inset in each panel plots the histograms of \mbh; green
and red lines denote the \mgii-based masses from our Schemes 2 and 4, and
black lines are the comparison masses.
The top three panels compare our \mgii\ masses with masses derived from
different \hb\ formalisms.  The \hb\ masses from Vestergaard \& Peterson
(2006; a) are systematically different from our \mgii\ masses, while
those of Collin et al. (2006; b) are roughly consistent, and the best
agreement comes from our newly derived formalism (Equation (9); c, red
points).  The bottom three panels compare our \mgii-based masses with previous
\mgii-based formalisms: McLure \& Dunlop (2004; d), Kollmeier et al.
(2006; e), and Salviander et al. (2007; f). All show systematic
deviations, mostly in the sense of giving lower masses than our formalism.}
\end{figure}
\clearpage

%\LongTables
\begin{deluxetable}{lrcccccc}
\centering
\tablenum{1}
\tabletypesize{\scriptsize}
\tablecaption{Data for the Reverberation-mapped Sample} \tablehead{\colhead{Name} &
\colhead{DataID} & \colhead{$\log L_{3000}$} & \colhead{ FWHM(Mg\,II)} &
\colhead{$\log L_{5100}$} & \colhead{ FWHM(\hb, mean)} & \colhead{ $\sigma_{line}$(\hb, rms)} & \colhead{$\log M_{\rm BH}({\rm RM})$} \\
\colhead{ } & \colhead{ } & \colhead{(\lum)} & \colhead{(km~s$^{-1}$)} & \colhead{(\lum)} &
\colhead{(km~s$^{-1}$)} & \colhead{(km~s$^{-1}$) } & \colhead{($10^6 M_\odot$)} \\
\colhead{(1)}  & \colhead{(2)} & \colhead{(3)} & \colhead{(4)} &
\colhead{(5)} & \colhead{(6)} & \colhead{(7)} & \colhead{(8) }}
\startdata
3C\,120         & lwp04153 &   44.46$\pm$  0.04 &    2780 &                &             &             &       \\
                & lwp04500 &   44.37$\pm$  0.03 &    3837 &                &             &             &       \\
                & lwp05610 &   44.35$\pm$  0.02 &    2568 &                &             &             &       \\
                & lwp09048 &   44.41$\pm$  0.03 &    2638 &                &             &             &       \\
                & lwp09461 &   44.20$\pm$  0.02 &    3411 &                &             &             &       \\
                & lwp09850 &   44.22$\pm$  0.02 &    3399 &                &             &             &       \\
                & lwp10407 &   44.29$\pm$  0.04 &    4083 &                &             &             &       \\
                & lwp11524 &   44.28$\pm$  0.03 &    2828 &                &             &             &       \\
                & lwp11946 &   44.27$\pm$  0.03 &    2786 &                &             &             &       \\
                & lwp12536 &   44.36$\pm$  0.03 &    2412 &                &             &             &       \\
                & lwr01317 &   43.86$\pm$  0.08 &    2874 &                &             &             &       \\
                & lwr02983 &   43.73$\pm$  0.02 &    2759 &                &             &             &       \\
                & lwr06849 &   44.12$\pm$  0.03 &    3997 &                &             &             &       \\
                & lwr09102 &   44.34$\pm$  0.03 &    2875 &                &             &             &       \\
                & lwr09778 &   43.93$\pm$  0.02 &    2735 &                &             &             &       \\
                & lwr13786 &   44.30$\pm$  0.02 &    2758 &                &             &             &       \\
                & lwr15618 &   44.25$\pm$  0.03 &    3040 &                &             &             &       \\
                & lwr16609 &   44.17$\pm$  0.04 &    3053 &                &             &             &       \\
                & lwr16874 &   44.39$\pm$  0.06 &    4431 &                &             &             &       \\
                &          &\bf{44.23$\pm$ 0.01}&\bf{3074}& 44.09$\pm$0.09 & 2327$\pm$50 & 1166$\pm$50 & $55.5^{+31.4}_{-22.5}$ \\
3C\,390.3       & y33y0204t&   42.63$\pm$  0.17 &    7884 & 43.64$\pm$0.14 &12694$\pm$13 & 3105$\pm$81 & $287\pm64$ \\
Akn\,120        & y29e0305t&   44.48$\pm$  0.08 &    4377 & 43.93$\pm$0.04 & 6143$\pm$42 & 1921$\pm$60 & $150\pm19$ \\
Fairall\,9      & y0ya0104t&   44.30$\pm$  0.08 &    3769 & 43.94$\pm$0.10 & 5999$\pm$66 & 3787$\pm$196& $255\pm56$ \\
IC4329A         &          &                    &         & 42.89$\pm$0.15 & 5964$\pm$134& 2476$\pm$226& $9.90^{+17.88}_{-11.88}$ \\
Mrk\,79         & lwr01320 &   43.81$\pm$  0.02 &    5057 &                &             &             &       \\
                & lwr06141 &   43.60$\pm$  0.02 &    4179 &                &             &             &       \\
                &          &\bf{43.71$\pm$ 0.02}&\bf{4597}& 43.65$\pm$0.03 & 4858$\pm$38 & 1882$\pm$121& $52.4\pm14.4$ \\
Mrk\,110        & lwp12760 &   43.64$\pm$  0.08 &    2216 &                &             &             &       \\
                & lwp12761 &   43.82$\pm$  0.07 &    2504 &                &             &             &       \\
                &          &\bf{43.73$\pm$ 0.05}&\bf{2355}& 43.66$\pm$0.04 & 1600$\pm$13 & 1002$\pm$53 & $25.1\pm6.1$ \\
Mrk\,279        & lwp02522 &   44.03$\pm$  0.03 &    4812 &                &             &             &       \\
                & lwp10116 &   43.77$\pm$  0.02 &    3314 &                &             &             &       \\
                & lwp15450 &   43.03$\pm$  0.04 &    5330 &                &             &             &       \\
                & lwp15687 &   43.75$\pm$  0.02 &    3275 &                &             &             &       \\
                & lwp19173 &   43.98$\pm$  0.04 &    3555 &                &             &             &       \\
                & lwp19220 &   44.02$\pm$  0.03 &    3987 &                &             &             &       \\
                & lwp19598 &   44.28$\pm$  0.05 &    3699 &                &             &             &       \\
                & lwp19937 &   44.20$\pm$  0.04 &    4204 &                &             &             &       \\
                & lwp20271 &   44.26$\pm$  0.05 &    4305 &                &             &             &       \\
                & lwp20725 &   44.18$\pm$  0.03 &    5057 &                &             &             &       \\
                & lwr03073 &   43.03$\pm$  0.10 &    3710 &                &             &             &       \\
                & lwr10816 &   43.98$\pm$  0.04 &    8201 &                &             &             &       \\
                & lwr11623 &   43.94$\pm$  0.02 &    6934 &                &             &             &       \\
                & lwr15803 &   43.96$\pm$  0.03 &    4121 &                &             &             &       \\
                &          &\bf{43.89$\pm$ 0.01}&\bf{4441}& 43.66$\pm$0.08 & 5354$\pm$32 & 1420$\pm$96 & $34.9\pm9.2$ \\
Mrk\,335        & y29e0205t&   44.12$\pm$  0.04 &    1977 &                &             &             &       \\
                & y29e0206t&   44.13$\pm$  0.05 &    1783 &                &             &             &       \\
                &          &\bf{44.13$\pm$ 0.03}&\bf{1878}& 43.78$\pm$0.02 & 1735$\pm$2  &  933$\pm$64 & $14.2\pm3.7$ \\
Mrk\,509        & y0ya0305t&   44.55$\pm$  0.05 &    3357 & 44.16$\pm$0.10 & 3015$\pm$2  & 1276$\pm$28 & $143\pm12$ \\
Mrk\,590        &          &                    &         & 43.46$\pm$0.05 & 2906$\pm$89 & 1231$\pm$49 & $47.5\pm7.4$ \\
Mrk\,817        & lwr11936 &   44.07$\pm$  0.03 &    3597 &                &             &             &       \\
                & lwr13704 &   44.03$\pm$  0.04 &    4565 &                &             &             &       \\
                &          &\bf{44.05$\pm$ 0.03}&\bf{4053}& 43.64$\pm$0.14 & 4899$\pm$37 & 1680$\pm$67 & $49.4\pm7.7$ \\
NGC\,3227       & o5kp01010&   41.76$\pm$  0.04 &    3688 & 42.48$\pm$0.04 & 5103$\pm$160& 1925$\pm$124& $42.2\pm21.4$ \\
NGC\,3516       & y31r0105t&   43.05$\pm$  0.02 &    3919 &                &             &             &       \\
                & y31r0206t&   43.23$\pm$  0.02 &    3775 &                &             &             &       \\
                & y31r0306t&   43.19$\pm$  0.02 &    3796 &                &             &             &       \\
                & y31r0406t&   42.73$\pm$  0.02 &    3499 &                &             &             &       \\
                & y31r0506t&   43.08$\pm$  0.02 &    3527 &                &             &             &       \\
                &          &\bf{43.06$\pm$ 0.01}&\bf{3699}& 42.62$\pm$0.28 &5840$\pm$1976& 1837$\pm$115& $42.7\pm14.6$ \\
NGC\,3783       & o57b01010&   43.39$\pm$  0.06 &    2524 & 43.02$\pm$0.06 & 3770$\pm$68 & 1753$\pm$141& $29.8\pm5.4$ \\
NGC\,4051       & lwp11100 &   41.76$\pm$  0.02 &    1574 &                &             &             &       \\
                & lwp12092 &   41.70$\pm$  0.02 &    1387 &                &             &             &       \\
                & lwp12092 &   41.68$\pm$  0.02 &    1503 &                &             &             &       \\
                & lwp19265 &   41.58$\pm$  0.02 &    1305 &                &             &             &       \\
                & lwp20497 &   41.53$\pm$  0.02 &     946 &                &             &             &       \\
                & lwp23153 &   41.70$\pm$  0.02 &     918 &                &             &             &       \\
                & lwp24347 &   41.81$\pm$  0.02 &     752 &                &             &             &       \\
                & lwp27297 &   41.77$\pm$  0.02 &    1349 &                &             &             &       \\
                & lwp27298 &   41.77$\pm$  0.02 &    1667 &                &             &             &       \\
                & lwr01728 &   41.76$\pm$  0.02 &    2582 &                &             &             &       \\
                &          &\bf{41.68$\pm$ 0.01}&\bf{1322}& 41.88$\pm$0.08 &  654$\pm$2  &  916$\pm$64 & $1.58^{+0.50}_{-0.65}$ \\
NGC\,4151       & o42303070&   43.04$\pm$  0.03 &    4905 &                &             &             &       \\
                & o59701040&   42.20$\pm$  0.02 &    3020 &                &             &             &       \\
                &          &\bf{42.62$\pm$ 0.02}&\bf{3849}& 41.92$\pm$0.23 & 6371$\pm$150& 1914$\pm$42 & $45.7^{+5.7}_{-4.7}$ \\
NGC\,4593       & lwp02731 &   42.82$\pm$  0.02 &    4123 &                &             &             &       \\
                & lwp05348 &   42.82$\pm$  0.02 &    3022 &                &             &             &       \\
                & lwp05371 &   42.81$\pm$  0.02 &    3288 &                &             &             &       \\
                & lwp05394 &   42.83$\pm$  0.02 &    2778 &                &             &             &       \\
                & lwp05411 &   42.80$\pm$  0.02 &    3444 &                &             &             &       \\
                & lwp05430 &   42.74$\pm$  0.02 &    3107 &                &             &             &       \\
                & lwp06266 &   42.64$\pm$  0.02 &    3237 &                &             &             &       \\
                & lwp06300 &   42.72$\pm$  0.02 &    3033 &                &             &             &       \\
                & lwp12278 &   43.01$\pm$  0.02 &    2822 &                &             &             &       \\
                & lwp12279 &   43.03$\pm$  0.02 &    5194 &                &             &             &       \\
                & lwr07884 &   42.83$\pm$  0.02 &    2723 &                &             &             &       \\
                & lwr09818 &   42.77$\pm$  0.02 &    1929 &                &             &             &       \\
                & lwr10539 &   42.86$\pm$  0.02 &    2987 &                &             &             &       \\
                & lwr10622 &   42.81$\pm$  0.02 &    3117 &                &             &             &       \\
                & lwr16177 &   42.87$\pm$  0.02 &    3321 &                &             &             &       \\
                &          &\bf{42.82$\pm$ 0.01}&\bf{3140}& 42.85$\pm$0.04 & 5143$\pm$16 & 1561$\pm$55 & $9.8\pm2.1$ \\
NGC\,5548       & y0ya0205t&   43.01$\pm$  0.03 &    4756 & 43.31$\pm$0.02 & 6107$\pm$23 & 2063$\pm$32 & $65.4^{+2.6}_{-2.5}$ \\
NGC\,7469       & y3b6010bt&   43.67$\pm$  0.03 &    3061 & 43.30$\pm$0.05 & 1722$\pm$30 & 1456$\pm$207& $12.2\pm1.4$ \\
PG\,0026$+$129  & y2jk0108t&   45.18$\pm$  0.04 &    1104 & 44.95$\pm$0.08 & 2544$\pm$56 & 1774$\pm$285& $393\pm96$ \\
PG\,0052$+$251  &          &                    &         & 44.78$\pm$0.12 & 5008$\pm$73 & 1783$\pm$86 & $369\pm76$ \\
PG\,0804$+$761  & lwr13645 &   45.14$\pm$  0.02 &    5175 &                &             &             &       \\
                & lwr16666 &   44.93$\pm$  0.12 &    2533 &                &             &             &       \\
                &          &\bf{45.04$\pm$ 0.06}&\bf{3621}& 44.88$\pm$0.09 & 3053$\pm$38 & 1971$\pm$105& $693\pm83$ \\
PG\,0844$+$349  & y0p80105t&   44.53$\pm$  0.04 &    3045 & 44.19$\pm$0.07 & 2694$\pm$58 & 1448$\pm$79 & $92.4\pm38.1$ \\
PG\,0953$+$414  &          &                    &         & 45.15$\pm$0.07 & 3071$\pm$27 & 1306$\pm$144& $276\pm59$    \\
PG\,1211$+$143  & y0iz0403t&   44.81$\pm$  0.03 &    1610 &                &             &             &       \\
                & y0iz0404t&   44.82$\pm$  0.03 &    1642 &                &             &             &       \\
                &          &\bf{44.82$\pm$ 0.02}&\bf{1626}& 44.70$\pm$0.08 & 2012$\pm$37 & 1080$\pm$102& $146\pm44$ \\
PG\,1226$+$023  & y0g4020et&   46.08$\pm$  0.06 &    3420 &                &             &             &       \\
                & y0g4020ft&   45.98$\pm$  0.06 &    3105 &                &             &             &       \\
                & y0g4020ht&   46.20$\pm$  0.09 &    2900 &                &             &             &       \\
                & y0g4020jt&   46.24$\pm$  0.08 &    2839 &                &             &             &       \\
                & y0g4020lt&   46.24$\pm$  0.07 &    3032 &                &             &             &       \\
                & y0g4020nt&   46.24$\pm$  0.07 &    3022 &                &             &             &       \\
                & y0nb0104t&   45.94$\pm$  0.06 &    3422 &                &             &             &       \\
                &          &\bf{46.13$\pm$ 0.03}&\bf{3098}& 45.93$\pm$0.06 & 3509$\pm$36 & 1777$\pm$150& $886\pm187$ \\
PG\,1229$+$204  & lwr13136 &   44.41$\pm$  0.07 &    3054 &                &             &             &       \\
                & lwr16071 &   44.47$\pm$  0.03 &    4940 &                &             &             &       \\
                &          &\bf{44.44$\pm$ 0.04}&\bf{3884}& 43.65$\pm$0.06 & 3828$\pm$54 & 1385$\pm$111& $73.2\pm35.2$ \\
PG\,1307$+$085  &          &                    &         & 44.82$\pm$0.05 & 5059$\pm$133& 1820$\pm$122& $440\pm123$ \\
PG\,1411$+$442  & o65617010&   44.81$\pm$  0.03 &    2452 & 44.52$\pm$0.05 & 2801$\pm$43 & 1607$\pm$168& $443\pm146$ \\
PG\,1426$+$015  & lwp05440 &   45.14$\pm$  0.03 &    6957 &                &             &             &       \\
                & lwp05446 &   45.04$\pm$  0.06 &    4575 &                &             &             &       \\
                & lwr16020 &   44.92$\pm$  0.07 &    6776 &                &             &             &       \\
                &          &\bf{45.04$\pm$ 0.03}&\bf{5997}& 44.60$\pm$0.09 & 7113$\pm$160& 3442$\pm$308& $1298\pm385$ \\
PG\,1613$+$658  & lwp19372 &   45.09$\pm$  0.03 &    7518 &                &             &             &       \\
                & lwp19380 &   45.10$\pm$  0.03 &    7996 &                &             &             &       \\
                &          &\bf{45.09$\pm$ 0.02}&\bf{7753}& 44.73$\pm$0.10 & 9074$\pm$103& 2547$\pm$342& $279\pm129$ \\
PG\,1617$+$175  & lwp07592 &   44.81$\pm$  0.06 &    5951 &                &             &             &       \\
                & lwp25629 &   44.64$\pm$  0.05 &    4375 &                &             &             &       \\
                &          &\bf{44.73$\pm$ 0.04}&\bf{5102}& 44.36$\pm$0.10 & 6641$\pm$190& 2626$\pm$211& $594\pm138$ \\
PG\,1700$+$518  &          &                    &         & 45.56$\pm$0.03 & 2252$\pm$85 & 1700$\pm$123& $781^{+182}_{-165}$     \\
PG\,2130$+$099  & lwp02520 &   44.60$\pm$  0.05 &    3042 &                &             &             &       \\
                & lwp03568 &   44.66$\pm$  0.08 &    2208 &                &             &             &       \\
                & lwp07205 &   44.53$\pm$  0.03 &    2039 &                &             &             &       \\
                & lwr01774 &   44.47$\pm$  0.06 &    1820 &                &             &             &       \\
                & lwr04610 &   44.52$\pm$  0.03 &    1290 &                &             &             &       \\
                & lwr04628 &   44.58$\pm$  0.06 &    2479 &                &             &             &       \\
                & lwr15802 &   44.51$\pm$  0.03 &    2885 &                &             &             &       \\
                &          &\bf{44.55$\pm$ 0.02}&\bf{2174}& 44.40$\pm$0.02 & 2853$\pm$39 & 1624$\pm$86 & $38\pm15$ \\
\enddata
\tablecomments{
Column (1) object name;
Column (2) identification name of the spectrum in {\it IUE} (prefix ``lw'')
or {\it HST}\ archives;
Column (3) monochromatic continuum luminosity at 3000 \AA. For objects with multiple spectra, mean monochromatic continuum luminosity at 3000 \AA\
is listed in bold at the bottom of each object.
Column (4) FWHM of broad Mg\,II. For objects with multiple spectra, mean FWHM is listed in bold at the bottom of each object.
Column (5) monochromatic continuum luminosity at 5100 \AA\ taken from Bentz et al. (2009). For objects having multiple measurements,
here listed is mean value.
Column (6) \hb\ FWHM of mean spectra taken from Collin et al. (2006) and recent update listed in Section 3.2. For objects having multiple measurements,
here listed is mean value.
Column (7) \hb\ $\sigma_{line}$ of rms spectra taken from Peterson et al. (2004) and recent update listed in Section 3.2. For objects having
multiple measurements, here listed is mean value.
Column (8) BH mass from RM taken from Peterson et al. (2004) and recent update listed in Section 3.2.}
\end{deluxetable}

\clearpage

\begin{deluxetable}{lccccccccc}
\centering
\tabletypesize{\tiny}
\tablenum{2}
\tablecaption{Continuum and Emission-line Parameters of the SDSS Sample}
\tablehead{
\colhead{SDSS Name} &
\colhead{$z$} &
\colhead{$\log L_{5100}$} &
\colhead{FWHM(H$\beta^b$)} &
\colhead{$\log F$(H$\beta^{b}$)} &
\colhead{$\log F$(H$\beta^{n}$)} &
\colhead{$\log L_{3000}$} &
\colhead{FWHM(Mg\,II$^b$)} &
\colhead{$\log F$(Mg\,II$^b$)} &
\colhead{$\log F$(Mg\,II$^n$)} \\
\colhead{(1)}  & \colhead{(2)} & \colhead{(3)} & \colhead{(4)} & \colhead{(5)}
& \colhead{(6)} & \colhead{(7)}  & \colhead{(8)} & \colhead{(9)} & \colhead{(10)}
}
\startdata
    J000011.96$+$000225.3  &  0.478  &  44.69  &   3037  & $-$13.89  & $-$15.68  &  44.99  &   3284  & $-$13.84  & $-$14.89  \\
    J000110.97$-$105247.5  &  0.529  &  44.98  &   6807  & $-$13.73  & $-$15.53  &  45.18  &   5797  & $-$13.85  & $-$15.15  \\
    J001725.36$+$141132.6  &  0.514  &  45.23  &   5676  & $-$13.49  & $-$15.52  &  45.49  &   4432  & $-$13.64  & $-$14.71  \\
    J002019.22$-$110609.2  &  0.492  &  44.85  &   2832  & $-$13.74  &   \nodata &  45.00  &   2677  & $-$13.91  & $-$16.16  \\
    J005121.25$+$004521.5  &  0.727  &  45.04  &   2572  & $-$14.09  & $-$15.07  &  45.14  &   1606  & $-$14.42  & $-$15.41  \\
    J005441.19$+$000110.7  &  0.646  &  45.08  &   2220  & $-$14.24  &   \nodata &  45.13  &   2172  & $-$14.20  & $-$15.29  \\
    J010448.57$-$091013.0  &  0.469  &  44.77  &   4610  & $-$13.81  & $-$15.33  &  44.88  &   2627  & $-$14.11  & $-$15.86  \\
    J010644.16$-$103410.6  &  0.468  &  44.72  &   3873  & $-$13.83  & $-$15.35  &  44.85  &   3074  & $-$13.80  & $-$15.43  \\
    J011132.34$+$133519.0  &  0.576  &  45.13  &   8060  & $-$13.66  & $-$15.72  &  45.39  &   5495  & $-$13.73  & $-$14.87  \\
    J012016.73$-$092028.8  &  0.495  &  44.71  &   3284  & $-$13.72  & $-$16.02  &  45.05  &   3312  & $-$13.58  & $-$16.07  \\
\enddata
\tablecomments{
Column (1) object name;
Column (2) redshift derived from the peak of [O\,III]~$\lambda 5007$.
Column (3) luminosity of the power-law continuum at 5100 \AA.
Column (4) FWHM of broad H$\beta$.
Column (5) flux of the broad component of \hb.
Column (6) flux of the narrow component of \hb.
Column (7) luminosity of the power-law continuum at 3000 \AA.
Column (8) FWHM of broad Mg\,II.
Column (9) flux of the broad component of Mg\, II.
Column (10) flux of the narrow component of Mg\, II.
Luminosities, fluxes, and FWHM are in units of \lum, \flux, and \kms,
respectively.(This table is available in its entirety in a machine-readable form in the online
journal. A portion is shown here for guidance regarding its form and content.)}
\end{deluxetable}

\clearpage

%%%%%%%%%%%%%%%%%%%%%%% table 3
\begin{deluxetable}{llrr}
\centering
%\tablewidth{15cm}
\tablenum{3} %\tabletypesize{\tiny}
%\rotate \tabletypesize{\tiny} %\tabletypesize{\scriptsize}
\tablecaption{Regression Results for $$\log \left[\frac{\rm FWHM(Mg\,II)}{1000~\kms}\right] = k \log \left[\frac{\rm FWHM(\hb)}{1000~\kms}\right] + c$$ }
\tablehead{\colhead{Method} && \colhead{$k$} & \colhead{$c$}}
\startdata
OLS             && $0.73\pm 0.02$  & $0.09\pm 0.01$  \\
WLS             && $0.73\pm 0.02$  & $0.09\pm 0.01$  \\
FITexy          && $0.77\pm 0.01$  & $0.06\pm 0.02$  \\
FITexy$\_$T02   && $0.81\pm 0.03$  & $0.04\pm 0.02$  \\
Gaussfit        && $0.81\pm 0.03$  & $0.04\pm 0.02$  \\
BCES (bisector)  && $0.78\pm 0.03$  & $0.06\pm 0.02$  \\
BCES (orthogonal)&& $0.81\pm 0.02$  & $0.05\pm 0.01$  \\
LINMIX$\_$ERR   && $0.79\pm 0.03$  & $0.05\pm 0.02$  \\
\enddata
\end{deluxetable}

%\clearpage

\begin{deluxetable}{lccccccccccc}
\tablenum{4} \centering \tabletypesize{\scriptsize}
%\rotate \tabletypesize{\tiny} %\tabletypesize{\scriptsize}
\tablecaption{Regression Results for $\log \left[\frac{M_{\rm BH}
{\rm (RM)}}{10^6 \, \msun}\right] = a + \beta
\log\left(\frac{L_{3000}}{10^{44}~\rm \lum}\right) + \gamma \log
\left[\frac{\rm FWHM(Mg\,II)}{1000~\kms}\right]$ and \mbh\
Comparisons} \tablehead{\colhead{Scheme} &  \colhead{$a$} &
\colhead{$\beta$} & \colhead{$\gamma$} & \multicolumn{2}{c}{$\Delta
M_{\rm BH}$(RM)}& \multicolumn{2}{c}{$\Delta M_{\rm BH}$(\hb)}&
\multicolumn{2}{c}{$\Delta M_{\rm BH}$(\hb)}&
\multicolumn{2}{c}{$\Delta M_{\rm BH}$(\hb)}\\
\colhead{} & \colhead{} & \colhead{} & \colhead{} &
\multicolumn{2}{c}{} & \multicolumn{2}{c}{(Vestergaard+06)} & \multicolumn{2}{c}{(Collin+06)} &
\multicolumn{2}{c}{(Ours)} \\
\colhead{(1)} & \colhead{(2)} & \colhead{(3)} & \colhead{(4)} &
\multicolumn{2}{c}{(5)} & \multicolumn{2}{c}{(6)} &
\multicolumn{2}{c}{(7)} & \multicolumn{2}{c}{(8)} \\
\cline{5-6} \cline{7-8} \cline{9-10} \cline{11-12}
\colhead{} & \colhead{} & \colhead{} & \colhead{} &
\colhead{Mean} & \colhead{1 $\sigma$} & \colhead{Mean} & \colhead{1 $\sigma$} &
\colhead{Mean} & \colhead{1 $\sigma$} & \colhead{Mean} &
\colhead{1 $\sigma$}} \startdata
Scheme 1\tablenotemark{a}  & $1.15\pm 0.27$  & $0.46\pm 0.08$ & $1.48\pm0.49$ & 0.01 & 0.39 & 0.15 & 0.22 & 0.10 & 0.14  & 0.07 & 0.12  \\
Scheme 2                   & $0.88\pm 0.08$  & $0.48\pm 0.08$ &    2          & 0.01 & 0.42 & 0.11 & 0.19 & 0.06 & 0.18  & 0.03 & 0.18  \\
Scheme 3                   & $1.03\pm 0.08$  & $0.48\pm 0.08$ &    1.70       & 0.01 & 0.40 & 0.13 & 0.20 & 0.07 & 0.15  & 0.05 & 0.14  \\
Scheme 4                   & $1.13\pm 0.27$  &      0.5       & $1.51\pm0.49$ & 0.01 & 0.40 & 0.10 & 0.21 & 0.05 & 0.14  & 0.03 & 0.12  \\
\enddata
\tablecomments{Fits for the 29 objects with Mg\,II
data measured in the paper. $\Delta M_{\rm BH} \equiv \log M_{\rm
BH} - \log M_{\rm BH}({\rm Mg\,II})$ are the differences of the
masses obtained from different methods with masses estimated by our
Mg\,II formalism for each scheme. Column (5) The mean  and
 standard deviation of the deviations between
BH masses obtained from RM and our single-epoch Mg\,II estimators.
The mean  and standard deviation of the deviations between
 masses estimated from our single-epoch Mg\,II estimator and masses
derived using
(Column 6) the \hb\ formalism of Vestergaard \& Peterson (2006),
(Column 7) the \hb\ formalism of Collin et al. (2006), and
(Column 8) the new \hb\ formalism obtained in this work (Equation (11)) for the SDSS sample.}
\tablenotetext{a}{Scheme 1 is fitted by using the code MLINMIX$\_$ERR of Kelly (2007).}
\end{deluxetable}
\end{document}